# STM observation of a box-shaped graphene nanostructure appeared after mechanical cleavage of pyrolytic graphite

**Rostislav V. Lapshin**[a, b]

[a]*Solid Nanotechnology Laboratory, Institute of Physical Problems, Zelenograd, Moscow, 124460, Russian Federation*

[b]*Department of Photosensitive Nano and Microsystems, Moscow Institute of Electronic Technology, Zelenograd, Moscow, 124498, Russian Federation*

E-mail: rlapshin@gmail.com

**Abstract** A description is given of a three-dimensional box-shaped graphene (BSG) nanostructure formed/uncovered by mechanical cleavage of highly oriented pyrolytic graphite (HOPG). The discovered nanostructure is a multilayer system of parallel hollow channels located along the surface and having quadrangular cross-section. The thickness of the channel walls/facets is approximately equal to 1 nm. The typical width of channel facets makes about 25 nm, the channel length is 390 nm and more. The investigation of the found nanostructure by means of a scanning tunneling microscope (STM) allows us to draw a conclusion that it is possible to make spatial constructions of graphene similar to the discovered one by mechanical compression, bending, splitting, and shifting graphite surface layers. The distinctive features of such constructions are the following: simplicity of the preparation method, small contact area between graphene planes and a substrate, large surface area, nanometer cross-sectional sizes of the channels, large aspect ratio. Potential fields of application include: ultra-sensitive detectors, high-performance catalytic cells, nanochannels for DNA manipulation, nanomechanical resonators, electron multiplication channels, high-capacity sorbents for hydrogen storage.

## 1. Introduction

From the moment of graphene discovery and until the present time, several methods of its preparation have been suggested [1, 2, 3, 4, 5]. Among the suggested methods, the method of mechanical exfoliation of graphene planes from highly oriented pyrolytic graphite (HOPG) [1, 5] deserves a special mention since mechanical exfoliation of graphene planes apparently underlies mechanism of formation of the spatial box-shaped graphene (BSG) nanostructure described in the present work.

A surface of HOPG having unusual appearance is presented in Fig. 1 [6]. The surface has been either formed or uncovered after mechanical cleavage. As a rule, plane atomically smooth areas with sizes from several hundreds of nanometers to several microns are produced after cleaving this sort of graphite [7]. In the case considered, the graphite surface represents a multilayer system of parallel hollow channels which plane facets/walls are apparently graphene sheets.

A periodical microstructure that appeared after mechanical cleavage of HOPG is described in work [8]. The microstructure is a system of parallel folds periodically repeating through approximately 100 μm. The width of a fold area makes about 2 μm. The microstructure consists of several graphite layers and reaches 1-2 μm in depth. The microstructure and the detected nanostructure have some similarities: both structures extend in one dimension,

# STM observation of a box-shaped graphene nanostructure

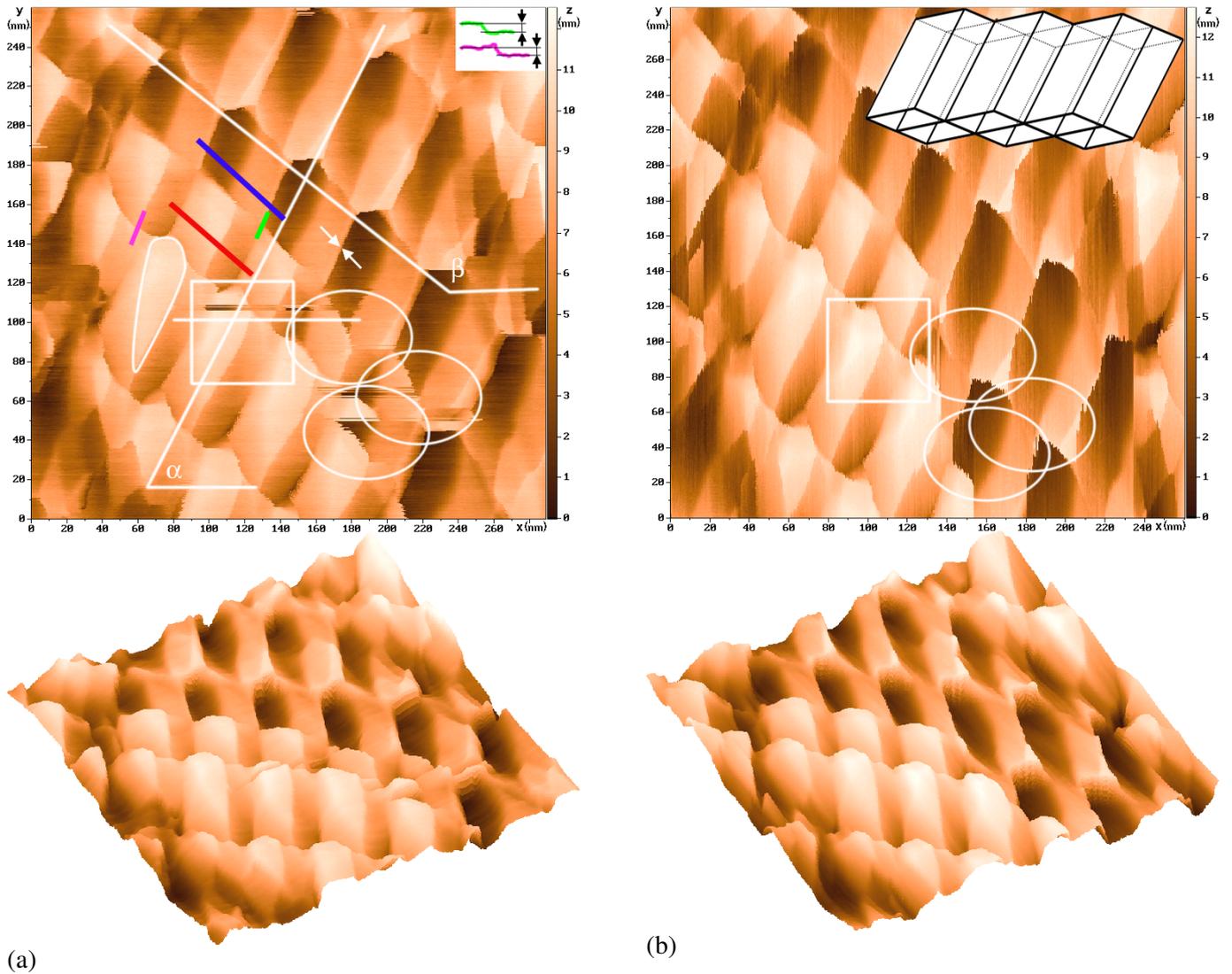

(a)  (b)

Fig. 1. 3D box-shaped nanostructure of graphene. The nanostructure represents a multilayer system of parallel hollow channels of a parallelogram shape in cross-section. A schematic view of the nanostructure is shown in the inset to figure part b. The wall/facet thickness (shown with white/black arrows) of the nanostructure channels makes about 1 nm. The 512×512 points STM-image is obtained in air in constant-current mode, $U_{tun}$=50 mV (sample positive), $I_{tun}$=890 pA. Fast-scan direction coincides with axis (a) $x$, (b) $y$. Structure regions inside the ovals became severely deformed as the direction of fast scanning had changed from $x$ to $y$. Channel orientation $\alpha$=62.7°, orientation of the facet cuts of the open channels $\beta$=143.8°.

periodically repeat, their folds are formed across the cleaving front, they both have layer shifts and channels with quadrangular cross-section. The observed similarities may imply similarity of the processes of formation of those surface structures, i. e., scalability of the phenomenon when passing from micrometer to nanometer fold sizes.

The main objectives of the presented work are

(1) Demonstration of the existence of BSG nanostructure.

(2) Analysis of the sizes and morphology of elements of BSG nanostructure.

(3) Development of a possible mechanism (a qualitative model) of the BSG nanostructure formation.

(4) A brief estimation of the prospects of possible applications of BSG nanostructure (to prove the need of further research).

Theoretical analysis and computer modeling of the discovered nanostructure as well as attempts of its reproduction are planned to be implemented at the next stages of the research. Based on the study of the BSG nanostructure, possible areas of its application were defined: detectors, catalytic cells, nanochannels of fluidic de-

R. V. Lapshin

vices, nanomechanical resonators, multiplication channels of electrons, hydrogen storage and some others.

The notions of a channel wall and a channel facet used below are close to each other. Wall, as a rule, refers to a flat surface common to two adjacent channels. Facets usually refer to outer flat surfaces of the upper channel layer.

**2. Specimen and measurement method**

HOPG (Research Institute of Graphite, Russia) with mosaic spread angle 0.8° (density 2.24 g/cm$^3$, purity 99.999%) was used as a specimen. The specimen was as thin as 0.3 mm strip of 2×4 mm. Electrical insulation adhesive tape (KLL, Taiwan) of polyvinylchloride 0.13 mm in thickness was used for cleavage. The images of the BSG nanostructure of 512×512 points were obtained with the scanning tunneling microscope (STM) Solver™ P4 (NT-MDT Co., Russia) in the air at room temperature, in the constant current mode. The bias voltage made 50 mV (sample positive), the tunneling current 890 pA. A mechanically cut ∅0.3 mm NiCr wire was used in the capacity of the tip. The typical noise level of the tunneling current in the course of the measurements made about 20 pA (peak-to-peak).

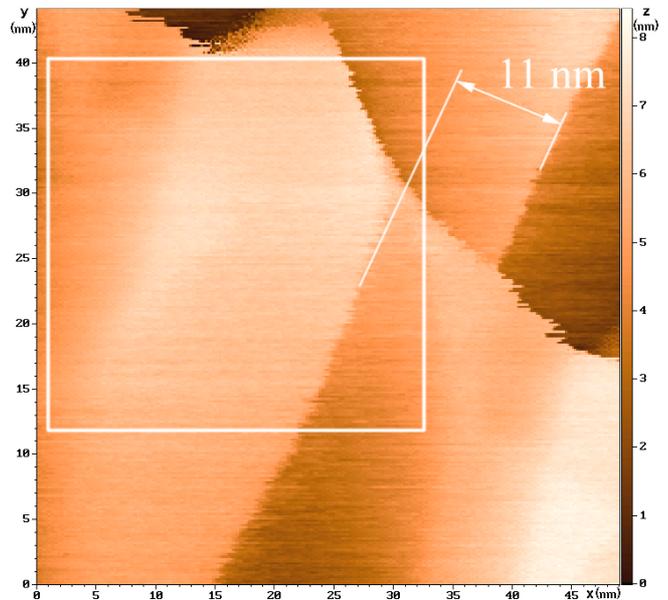

Fig. 2. Magnified image of the top facet of the box-shaped nanostructure channel. Thickness of the upper graphene layer makes about 1 nm (see the magenta cross-section at the similar location in Fig. 1(a)). Lateral shift of the upper graphene layer relative to the underlying layer makes 11 nm. Fast-scan direction is the *x* axis.

**3. Experimental observations**

The following experimental facts point out the small thickness of the walls/facets of the detected nanostructure. First, the direct measurement of the wall thickness (see the white arrows in Fig. 1(a)) of an "open" channel gives a size of order of 1 nm (an open channel is the one that has no top facets). Second, the direct measurement of the facet thickness (see the black arrows in the inset) also gives a size of order of 1 nm.

Third, during the raster scanning, the STM tip seems to cause plastic deformation of the box-shaped nanostructure for some regions even with tunneling currents <1 nA. The latter can only take place in case of thin enough facets/walls of the nanostructure. In particular, one of the possible signs of such deformation is a flattening of top facets of the box-shaped structure. The flattening looks like a notable decrease of the slope of a nanostructure facet. As an example, one of such places is outlined with a curvilinear contour in Fig. 1(a). Similar formations are well seen on the facets of the neighboring channels. The top facet of a channel inside the frame in Fig. 1 is shown in Fig. 2 at a higher magnification.

Moreover, the small thickness of the facets/walls is pointed out by the fact of damage (or plastic deformation) of several regions (outlined with ovals) of the box-shaped nanostructure. The damage happened after changing fast scanning direction by 90° (compare the regions outlined with ovals in (b) with the same regions in (a)). Most likely, the stiffness of the facets/walls in *y* direction turned out to be insufficient to withstand the force influence from the STM tip. Noteworthy is that during the initial scanning along the fast direction coincided with *x* axis, the region with the overhanging edge (bottom oval) did not break and did not bend plastically although notable scanning faults caused by its mechanical instability were registered.

# STM observation of a box-shaped graphene nanostructure

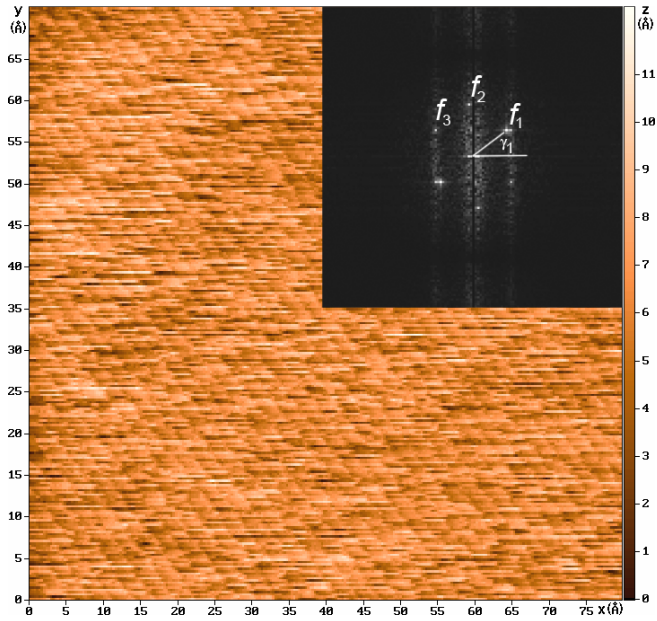

Fig. 3. Atomic resolution on a facet of a channel of the box-shaped nanostructure (constant current mode). Fast-scan direction is the $x$ axis. Fourier spectrum is given in the inset, where six maxima typical of graphite can be observed. The hexagon composed of the six maxima is significantly distorted by thermal drift and residual deformations of the lattice. The maxima of the spatial frequencies: $f_1=1/2.1$ Å$^{-1}$, $f_2=1/1.6$ Å$^{-1}$, $f_3=1/2.1$ Å$^{-1}$. The propagation directions of the spatial frequencies: $\gamma_1=41.1°$, $\gamma_2=94.7°$, $\gamma_3=148.0°$.

Let us remind that between the STM tip and the surface under investigation, besides a tunneling current registered during the STM measurement, a force interaction takes place [7, 9, 10]. Herein, the larger the tunneling current $I_{tun}$ (set point) is, the closer the microscope tip is located to the surface with the same bias voltage $U_{tun}$ applied and, in turn, the greater forces act between the tip apex and the surface.

Specific faults that appeared during the scanning are a fourth sign pointing out to a small thickness of the nanostructure facets/walls. These faults appear as narrow streaks from one to several raster lines in width (see the area located above the horizontal line in Fig. 1(a)). The streaks are oriented exactly along the raster lines. Such faults in microscope operation could be taken for a damage/modification of the surface resulted from the above mentioned forces acting between the STM tip and the surface since a damage/modification of the surface often leads to unstable scanning. However, the next scanning of the same surface areas revealed no signs of damage/modification and after switching the fast scanning direction from $x$ to $y$, the faults disappeared at all (see Fig. 1(b)).

The faults under consideration could also have been interpreted as a result of random surface pollutions. Such pollutions are often pollutions introduced from the outside or a nanoscale debris of the nanostructure originated from cleaving/scanning [7]. The nanoscale debris cause unstable microscope operation by falling under the probe and/or by sticking to it. The practice of STM measurements, however, shows that the presence of any pollutions, as a rule, would make scanning with atomic resolution either completely impossible or extremely unstable.

Meantime, a quite stable atomic resolution was obtained (see Fig. 3) at the subsequent scanning of the upper facet of the nanostructure with high magnification (small scanning step) near the middle of the area enclosed in a frame in Fig. 2. It is well seen in Fig. 2 that the area of the facet flattening, which had been previously considered as a single whole, under a higher magnification turned out to consist of several conditionally plane areas having slightly different slopes. The borders of the plane areas are composed of bent sections of graphene plane formed during a plastic deformation.

Taking into account the said above, the nature of the observed streaks can be as follows. If the scanned surface is the surface of a very thin membrane, then the forces applied by the STM tip while moving by the surface cause its elastic deformation (bending) [11, 12]. For example, under attractive van der Waals forces, the membrane bends toward the tip thus increasing the tunnel current. The microscope feedback loop immediately attempts to compensate the change in current by moving the scanner Z manipulator away from the surface. As a result, a not really existing increase in topography height will be observed on the obtained image.

Under the action of repulsive forces, the membrane, on the contrary, bends away from the tip. In this case, try-

**R. V. Lapshin**

ing to reach the set value of the tunneling current, the microscope feedback system moves the scanner's Z manipulator toward the surface thus increasing the membrane deformation even more. At a certain moment, the membrane reacting force is increased so much that it becomes equal to the tip pressing force and the tunneling current reaches the set value. As a result, a lowering of the topography height will be observed on the obtained image which is absent on the real surface. Taking into account the abrupt dependence of the tunneling current upon the size of tunnel junction, the pointed out topography changes can be rather strong.

It is just the described type of tip-surface force interaction that takes place in Fig. 1(a) in the form of the above fault. Abrupt topography falls (up to 6 nm in depth) and steep topography raisings (up to 4 nm in height) are clearly seen where they should not be judging by the adjacent scan lines.

It is worthy to note that the force interaction between an STM tip and a surface may possess a hysteresis. In that case, the force interaction has inelastic character [11] that may point out a rearrangement of the graphene lattice structure and/or a relative sliding of the graphene layers. In certain conditions, hysteresis of the force interaction can cause oscillations of the microscope feedback loop.

The analysis of the obtained STM images thus shows that the facets/walls of the channel of the box-shaped nanostructure are thin membranes (nanomembranes) with the typical thickness of 1 nm. Although the measured thickness of 1 nm corresponds to three graphene layers (the distance between neighboring graphite/graphene layers makes 3.35 Å [13]), the actual facet/wall thickness of the nanostructure can be a couple of graphene layers or even a single layer. This occurs because of some peculiarities of nanostructure wall formation (see splitting into sublayers in Section 5), widening of the objects due to interaction with a sidewall of the STM tip as well as deformation of the nanostructure itself in the locations where the thicknesses of the facets are being measured.

Because of small sizes of the elements of the box-shaped nanostructure, the membranes under consideration have characteristic oscillations of very high frequency [12, 14, 15, 16, 17]. As a result, these oscillations cannot pass through the low-frequency microscope servosystem. For the same reasons, the membranes cannot be excited by outside acoustic or seismic disturbances with frequencies entirely within the low-frequency spectral range.

It is well known that even as big objects as micromembranes and microcantilevers can get excited by thermal fluctuations already at room temperature [17, 18, 19, 20]. Besides, free oscillations of a nanomembrane can be excited/damped by the mechanical force exerted by the tip. For example, at the membrane edges (facet cuts of the open channels), high-frequency excitation can occur when the tip goes down from the membrane and, vice versa, it can be damped when the tip goes up onto the membrane. In both cases, the side surface of the tip would participate in the interaction. As a result, the root-mean-square level of noise would noticeably increase at the points corresponding to the edges of the membranes (see Figs. 1 and 2).

In Fig. 1(a), the hollow channels of the BSG nanostructure are oriented toward the scan axis $x$ at the angle $\alpha$=62.7° and the facet cuts of the open channels make the angle $\beta$=143.8° with the axis $x$. The appearance and the analysis of the cross-sections of the discovered channels have shown that the graphene facets/walls of the nanostructure are not perfect planes [21] and the form of the channel cross-section is close to a parallelogram (see Fig. 4). The large diagonal of the parallelogram is nearly parallel to the horizontal plane (basal plane). By the STM scans, the following was approximately determined: the mean channel depth $d$=8±1 nm, the mean size of width projection of the small channel facet $w_x$=18±1 nm, and the mean size of width projection of the large channel facet $W_x$=28±1 nm, the channel length $L$ makes 390 nm and more.

# STM observation of a box-shaped graphene nanostructure

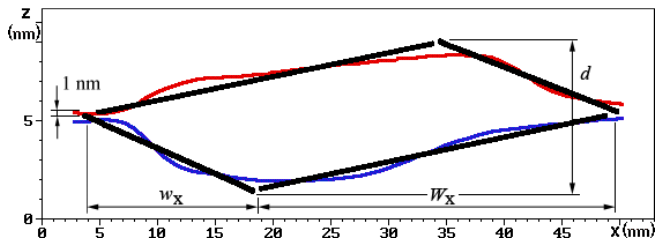

Fig. 4. Cross-sectional view of the "open" (low profile) and the "closed" (upper profile) parts of a channel of the box-shaped nanostructure. Matching of the profiles shows that the cross-section shape of the discovered nanostructure is close to a parallelogram. The section locations are shown in Fig. 1(a) with thick lines of corresponding colors. Mean channel depth $d=8\pm1$ nm, mean sizes of width projections of small $w_x=18\pm1$ nm and large $W_x=28\pm1$ nm facets of the channel.

## 4. Analysis of the observations

Since the nanostructure under consideration looks periodical, we may expect well noticeable maxima, corresponding to the observed periodicity, to be present at the two-dimensional Fourier spectrum of the nanostructure. Fig. 1(a) gives a good idea of the directions along which some possible periodicities could propagate. These directions are defined by $\alpha+90°=152.7°$ and $\beta-90°=53.8°$ angles. Along the first of the above directions, the channels themselves repeat periodically; along the second one – the cuts of the upper facets do, which are going nearly perpendicular to these channels. The angle $\alpha+90°$ apparently points out the direction of the cleaving front movement. The cleaving front refers to the moving line of a contact of the adhesive tape with a graphite surface.

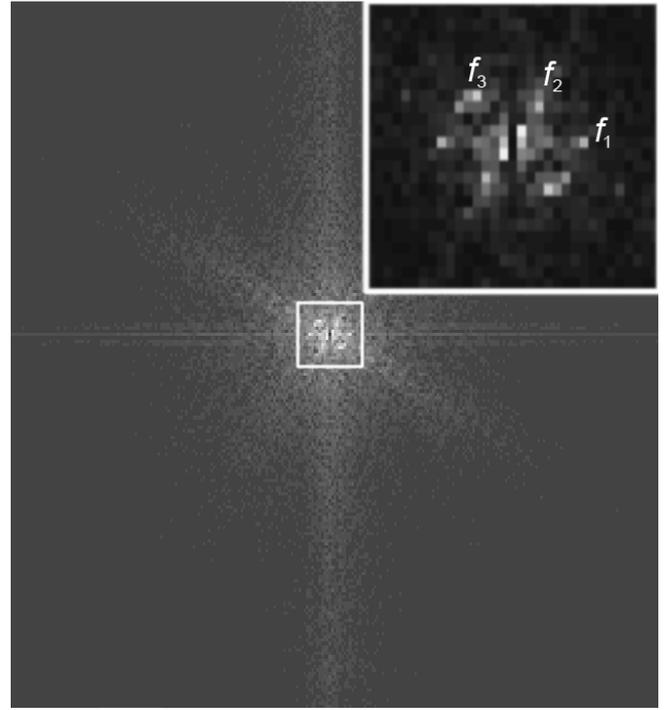

Fig. 5. Fourier spectrum of the box-shaped nanostructure. Spatial periods of the most pronounced surface topography oscillations: $f_1^{-1}=36.1$ nm, $f_2^{-1}=64.5$ nm, $f_3^{-1}=48.3$ nm. The directions of oscillation propagation corresponding to the found periods: $\gamma_1=0.0°$, $\gamma_2=48.1°$, $\gamma_3=139.5°$. For better visualization, the spectrum image is normalized in the vertical plane with the use of a nonlinear (logarithmic) scale.

Indeed, the expected oscillations are present in the Fourier spectrum (see Fig. 5). In particular, three maxima with the spatial periods $f_1^{-1}=36.1$ nm, $f_2^{-1}=64.5$ nm, and $f_3^{-1}=48.3$ nm are well distinguishable. The detected oscillations $f_1, f_2, f_3$ propagate at the angles $\gamma_1=0.0°$, $\gamma_2=48.1°$, and $\gamma_3=139.5°$ to the spectrum axis $x^{-1}$, respectively. As expected, the spatial period $f_3^{-1}$ is very close to the manually measured sum of the projections of the widths of the upper facets $w_x+W_x=46$ nm. However, the direction this oscillation propagates along differs from the expected direction by more than 13°. Since Fourier spectrum is capable of the best estimate of the spatial period mean value, more precise values of the width projections of the facets $w_x=18.9$ nm and $W_x=29.4$ nm are further used in calculations (to the mean values $w_x$ and $W_x$ measured manually, 0.9 nm and 1.4 nm were added, respectively, so as the total facet width $w_x+W_x$ would numerically equalized with the found period $f_3^{-1}$).

The spatial period $f_2^{-1}$ relates to the periodicity in the locations of the cross cuts of the upper facets; its direction $\gamma_2$ coincides quite well with the direction $\beta-90°$. The origination of the spatial period $f_1^{-1}$ is not that obvious, though. As this oscillation propagates strictly in the horizontal direction $\gamma_1$, some relation should be assumed between the oscillation and the movement along a raster line. However, the frequency $f_1$ having a comparable ampli-

R. V. Lapshin

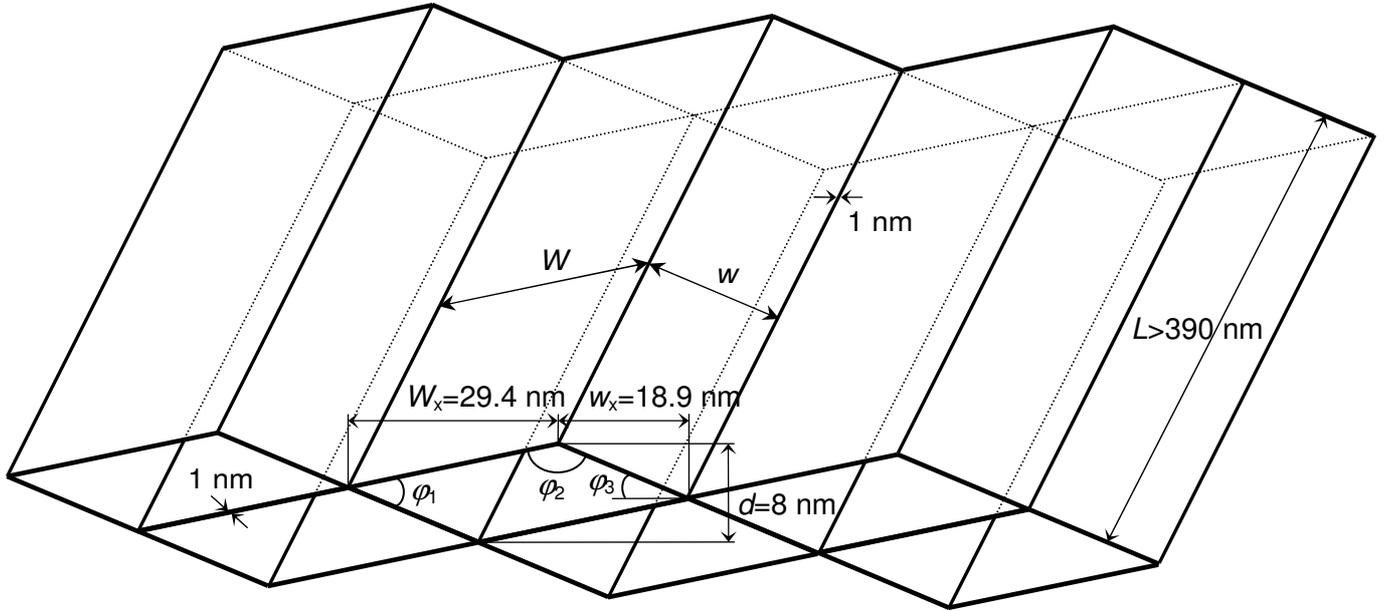

Fig. 6. A schematic representation of the box-shaped nanostructure and the typical dimensions of its elements. Thickness of the channel facets/walls makes about 1 nm. Width of the small facet $w$=19.3 nm, width of the large facet $W$=29.7 nm. The angles in the channel cross-section: $\varphi_1$=19.7°, $\varphi_2$=160.3°, $\varphi_3$=12.0°.

tude and oriented along $x^{-1}$ is also present on the Fourier spectrum built for the scan in Fig. 1(b), where the fast-scan direction coincides with $y$ direction.

By the obtained sizes $d$, $w_x$, and $W_x$, calculated were the widths of the small $w$=19.3 nm and the large $W$=29.7 nm facets ($w/W\approx 2/3$) as well as the parallelogram angles $\varphi_1$=19.7°, $\varphi_2$=160.3°, and $\varphi_3$=12.0°. The model representation of the BSG nanostructure and its typical sizes are given in Fig. 6.

It is rather difficult to suggest an unambiguous description of the mechanism responsible for the spatial box-shaped nanostructure formation based on the available data only. For example, it is still unclear whether the detected nanostructure was originated inside the HOPG body during its crystallization and then simply unsealed (uncovered) by the manual cleavage or whether this nanostructure was formed immediately during the mechanical cleaving in a surface layer. In case the nanostructure has formed immediately during the cleaving, whether the HOPG specimen had some structure peculiarities in the considered area by the moment of the nanostructure formation. Those peculiarities could be intercalations, ordered pattern of defects, *etc.* which had allowed for the observed nanostructure to be formed at the moment of the cleavage.

In the context of the above, noteworthy is the fact that the facets/walls that appeared on the surface make up the nanostructure almost completely consisting of plain areas. The prevalence of plain areas is another fact proving that graphene sheets are the main structural component of the considered nanostructure.

In the scientific literature there are many reports where images of complex dislocation networks observed with an STM on HOPG surface are presented [9, 22]. As a rule, dislocation networks observed with an STM are not registered with an atomic-force microscope (AFM). This fact means that the dislocation network is connected with some electronic properties of the HOPG sample and is physically located under the surface rather than upon it. In this regard, a question arises: whether the observed box-shaped nanostructure is really sort of a dislocation network.

Analysis of the published dislocation networks shows that the difference in height of the registered topography makes several Ångströms. In the observed nanostructure the difference in height makes 12-15 nm after elimination of a mean tilt and smoothing. Moreover, the appearance of the nanostructure (the shape of the elements, their mu-

# STM observation of a box-shaped graphene nanostructure

tual location and bulkiness) does not match any of the known substantially flat dislocation networks. Thus, according to the signs mentioned above, the detected box-shaped nanostructure may not be admitted as a dislocation network.

It is difficult to say anything definitive about the real number of the formed layers of the channels of the box-shaped nanostructure based only on the available data. At least, two layers of the channels are clearly recognizable in Fig. 1. The lowermost layer can be observed when moving along the diagonal connecting the left top and the right bottom corners of the scan (red and blue cutting lines). This layer is partially opened (blue cutting line). In parallel with this layer, yet another channel layer located above is well seen in the right top corner of the scan, which is partially opened as well. The two layers of the box-shaped channels are shown schematically in Fig. 6.

In parallel to the upper facets of the lower layer of the channels in the left bottom corner of the scan, two graphene layers are located (see Fig. 1). They overlap one another and have a thickness of about 1 nm each. Those graphene layers partially cover the upper facets of the lower layer of the channels. The edge of one of the graphene layers is shown in Fig. 2 with higher magnification. The fact of existence of these two clearly distinguishable graphene layers is yet another sign pointing out the small thickness of the facets/walls of the found nanostructure.

It is well seen in Fig. 2 that the upper graphene layer is shifted relative to the lower layer by approximately 11 nm in the lateral plane. Considering the profile of the nanostructure channels, it becomes obvious that an empty space should be formed between the layers after lateral shifting (see Section 5). The existence of the empty space and the force applied by the microscope tip to the upper graphene layer during scanning may explain why the plane upper facets of the channels were found somewhat deformed.

As it was noted above, atomic resolution is possible (see Fig. 3) in plain, almost horizontal regions of the facets (see Fig. 2), despite of heavily corrugated surface of the box-shaped nanostructure and small thickness of its walls/facets. Fourier spectrum of the surface is shown in the inset to Fig. 3, where six maxima of spatial frequencies typical of graphite/graphene are well distinguishable: $f_1$=1/2.1 Å$^{-1}$, $f_2$=1/1.6 Å$^{-1}$, and $f_3$=1/2.1 Å$^{-1}$. By the spatial frequencies $f$, the lattice constants $a_1$=2.1 Å, $a_2$=2.8 Å, $a_3$=2.1 Å were determined. The lattice constants $a_1$, $a_2$, $a_3$ differ noticeably (by 15%) from the lattice constant $a$=2.46 Å [13] of ordinary graphite.

The hexagon formed by the maxima is strongly distorted by thermal drift [23, 24, 25]. Moreover, the hexagon is probably distorted by residual deformations that appeared during formation of the structure and its scanning. In the absence of any distortions, the considered hexagon is regular.

In order to precisely determine, using an STM, the degree of residual strain in the lattice of graphene that the facets of the box-shaped nanostructure are formed of, the method of feature-oriented scanning (FOS) [23, 24, 25] should be applied. A distinctive feature of FOS is *in situ* elimination of drift influence on the scanning results. It is worth noting that atomic resolution was realy obtained on the surface of a thin membrane consisting of 2-3 graphene layers. This fact again confirms (inderectly) the high rigidity of graphene structures [1, 2, 3].

Propagation directions $\gamma_1$=41.1°, $\gamma_2$=94.7°, $\gamma_3$=148.0° of the spatial frequencies have also been determined by the Fourier spectra in Fig. 3. Given that those directions are known, it is easy to determine the crystallographic directions on the surface of the nanostructure facet: $\theta_1$=$\gamma_1$+90°=131.1°, $\theta_2$=$\gamma_2$-90°=4.7°, $\theta_3$=$\gamma_3$-90°=58.0° (see Fig. 7).

By comparing directions $\alpha$ and $\beta$ of the box-shaped nanostructure with the crystallographic directions $\theta$ on the facet surface, we can say with high degree of confidence that the direction $\alpha$ of the nanostructure channels ap-proximately coincides with the direction $\theta_3$ of the graphene plane and the direction $\beta$ of the facet cuts of the channels approximately coincides with the direction $\theta_1$ of the graphene plane [26]. Some discrepancies between the angles $\alpha$ and $\theta_3$, $\beta$ and $\theta_1$ can be accounted for by differences in drift velocities (thermal drift + creep) [24, 25], which probably took place as the scans shown in Fig. 1(a) and Fig. 3 were being acquired. Moreover, the fact of deformation of the lattice of the nanostructure facet should also be taken into account. An additional argument in favor of the drift influence is that the directions $\alpha$ and $\beta$ are deviated to the same side (counterclockwise) from the directions $\theta_3$ and $\theta_1$, respectively. Thus, to fabricate the box-shaped nanostructures under consideration, the cleaving front (see Fig. 7) should be oriented, at least approximately, along one of the graphite crystallographic directions $\theta$.

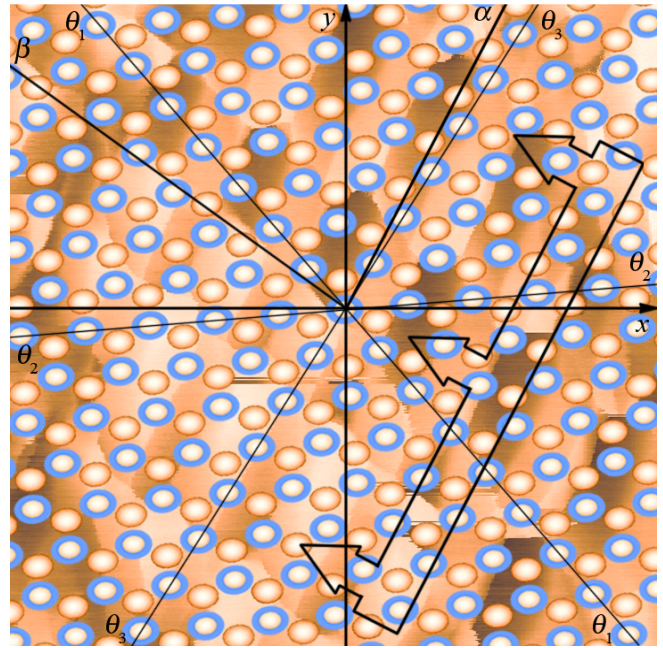

Fig. 7. Position of the box-shaped nanostructure relative to the initial crystal lattice of graphite. The rectangle with arrows indicates the probable direction of movement of the cleaving front. The carbon atoms usually observable with an STM are encircled with blue. The atomic surface is distorted by drifts and deformed.

Taking into consideration that the found values of lattice constants $a_1$, $a_2$, and $a_3$ are more like $a$=2.46 Å than like $a$=1.42 Å, we can suggest that the number of graphene layers in this particular facet is no less than two and that the relative location of the layers is exactly the same as the one of the adjacent layers in graphite (ABAB stacking).

**5. Formation mechanism**

Below is a qualitative description of the probable formation mechanism of the detected BSG nanostructure. It is assumed that the box-shaped nanostructure arises as a result of a mechanical cleaving performed by an adhesive tape. Fig. 8 shows the HOPG cleaving method that possibly enables the formation of the searched for box-shaped nanostructure. At first glance, the method might seem just insignificantly different from the existing one. Nevertheless, there are several specific peculiarities, namely:

(1) a small-valued (about 12°) cleavage angle $\varphi_3$ defined as the tilt of the small facet of the nanostructure channel (see Fig. 6);

(2) the position of the adhesive tape on the graphite surface so the cleavage front be approximately parallel to one of the crystallographic directions of the lattice (see Fig. 7);

(3) the setting of a minimal external cleaving force $F$ and keeping that force constant during the whole process.

Let us make a detailed analysis of the cleaving process. To begin with, let us consider some short-length (tens of nanometers) section AB of the cleaving surface directly adjacent to the current position of the cleavage front. The cleavage front passes through the point A normally to the plane of Fig. 8. The action of the external cleaving force $F$ is transferred through the adhesive tape to a thin surface layer of the considered section AB (pos. 1). Under the influence of a lateral component of the cleaving force, the graphite crystal lattice will undergo elastic compression at the section AB.

When this compression reaches a certain ultimate value, mechanical condition of the thin graphite surface

**STM observation of a box-shaped graphene nanostructure**

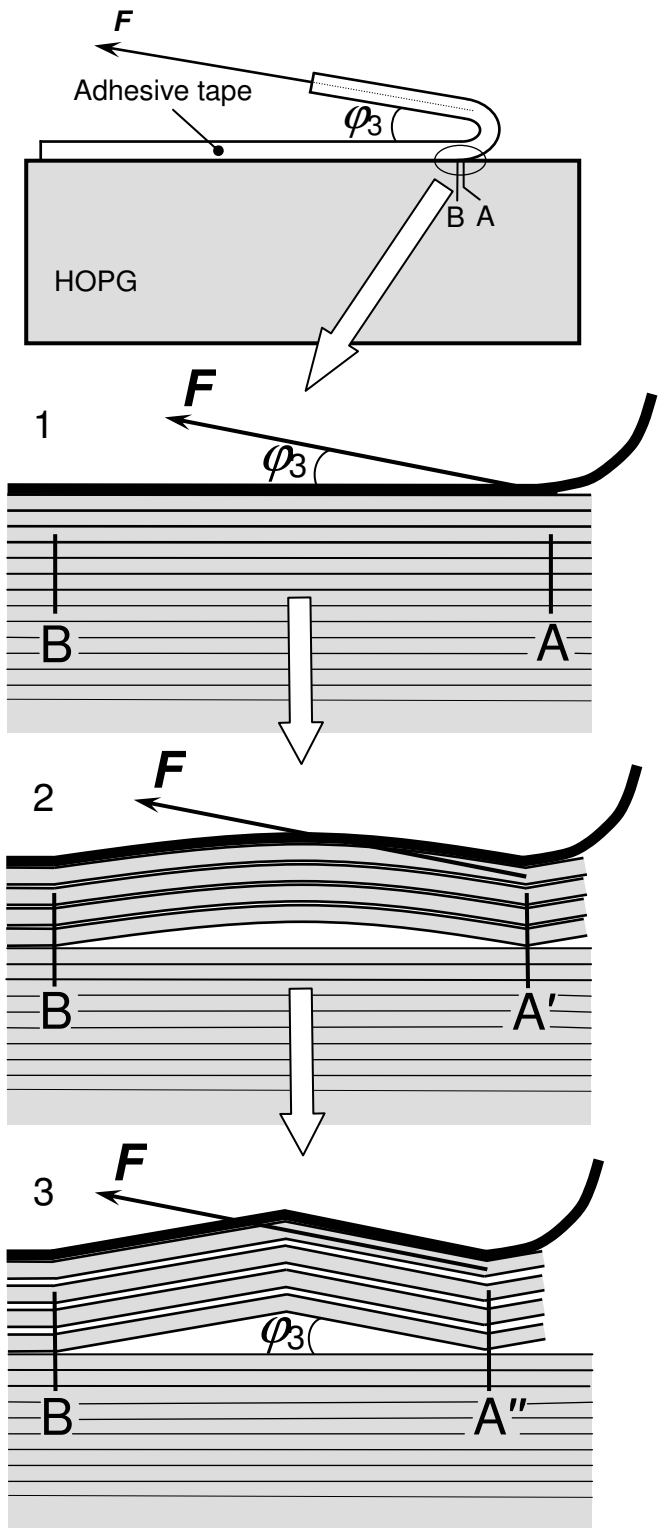

Fig. 8. Mechanism of nanofold formation: pos. 1 – elastic compression of a thin surface layer; pos. 2 – elastic bending of the thin surface layer and its detaching from the crystal; pos. 3 – plastic bending of the thin surface layer and its splitting into graphene sublayers. *F* stands for a cleaving force applied to the basal plane at the angle $\varphi_3$, AB=$w+W$, A″B=$w_x+W_x$. Proportions between certain elements are not met.

layer (this layer is conventionally represented by four graphene sublayers) becomes unstable at the section A′B and this layer starts elastically bending (pos. 2) and detaching from the main crystal body. The bending occurs along the solely possible direction – away from the surface – since the hardness of graphite is greater than the hardness of the adhesive layer of the tape.

As the bending proceeds, the cleaving front slides horizontally A→A′→A″ [27, 28]. Because of an abrupt decrease in attracting van der Waals forces acting at the section A′B between the considered graphite layer and the underlying crystal, the sliding speed of the cleaving front increases rapidly and plastic bending deformation occurs. In the course of the plastic deformation, a bend (fold) of the graphite layer is formed (pos. 3). During the plastic bending, the layer is also being split into several thinner (graphene) sublayers of nanometer thickness. The splitting into four graphene sublayers is conditionally shown in Fig. 8. The similar phenomenon, but observed on the microscale, was described in [8].

The fact that the observed graphite nanofolds are flat slops of a "roof" rather than a smoothly curving surface, could be explained by the atomic structure of the surface and by the particular orientation of the cleavage front relative to the crystallographic directions of the surface.

Taking into account the widths of the small $w$ and the large $W$ facets as well as their projections $w_x$ and $W_x$, it is easy to determine the shift $w+W-w_x-W_x=0.7$ nm of the cleaving front from point A to point A″ that resulted in formation of the observed fold. The actual shift of the front has most likely been somewhat greater since the obtained value corresponds to the state of the fold at the moment when the cleaving force was removed (the adhesive tape detached), *i. e.*, after a certain inevitable elastic fold relaxation. A reverse transformation of the fold into the original "stack" of graphene sheets does not occur since the formation of the fold was accompanied by the inelastic processes of bending, splitting into sub-layers and sliding the cleaving front from point A′ to point A″.

As the above process proceeds, the adhesive tape comes finally detached at the section A″B under the influence of the vertical component of the cleaving force *F*. Moreover, the tape along with a graphite layer stuck to it

R. V. Lapshin

detaches at the split boundary of the graphene layers in the fold. Further on, the above-described process of the fold formation repeats itself at a new section AB.

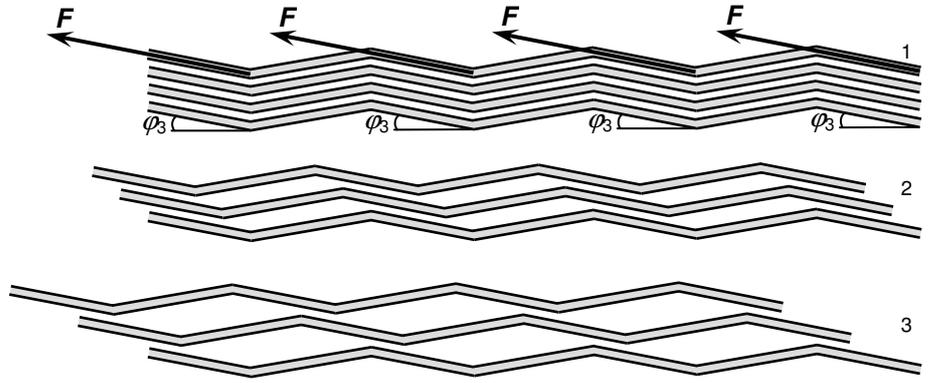

Fig. 9. Simplified formation mechanism of the channel layers of the box-shaped nanostructure (cross-sectional view). Two channel layers appear from three split-in-folds graphene layers during relative shifting (sliding) of these layers along the plane of the small facet under the action of a cleaving force $F$. $\varphi_3$ is the angle of force application, *i. e.*, the tilt of the small facet plane to the basal plane.

Now, let us show, within the framework of the suggested model of the nanofold formation, how graphene nanostructures consisting of one or more layers of hollow channels can appear. It was noted above that simultaneously with the formation of the folds, the graphite layer in them most likely splits into several thinner sublayers. Experimental confirmation of this assumption are Fig. 1 and Fig. 2 where such splitting can be directly observed. It is clearly seen in Fig. 1 that at least two split graphene layers exist on the upper facets of the nanochannels. The thickness of each layer makes about 1 nm.

Moreover, the layer located at the upper facets of the channel is shifted relative to these facets by approximately 11 nm in the direction defined by angle $\alpha+90°$. The layer located above the mentioned layer is also shifted relative to that layer by approximately the same value in the same direction. By the way, the movement direction of the cleavage front in Fig. 7 is chosen exactly based on this observation. The existing shift is undoubtedly a direct confirmation of the possibility that the split graphite layers can shift in a fold relative to each other at nanoscopic scale.

Thus, the presented facts point out that the graphene nanostructure consisting of one or more layers of channels can be formed as a result of a relative shift (sliding) of the split graphene layers in a fold under the action of cleaving force $F$. The angle the force $F$ is applied apparently should be determined by the angle $\varphi_3$ (see Fig. 6), *i. e.*, the slope of the small facet of the nanostructure channel to the horizontal plane (basal plane of graphite).

Fig. 9 schematically shows the formation of a two-layer BSG nanostructure during a relative shift of three split graphene layers in folds. For simplicity reasons, the cross-sections of the adjacent channels are shown in the figure as increasing simultaneously during the cleaving process (actually, the end result of the cleaving is presented). The real picture is different, though. In particular, the cross-sections of the channels would become larger as the cleaving front comes nearer. A single microchannel was observed in [8] that had a quadrangular cross-section and had been formed by shifting in a fold of graphite layers of submicron thickness.

The condition that the external cleaving force $F$ should be set to the minimum value is dictated by a relatively slow consecutive nature of the processes: compression-bending of the surface graphite layer, nanofold formation-splitting, relative shifts of the graphene layers in the nanofolds. The condition of keeping a constant value of the cleaving force $F$ during the entire process is to ensure that the elements of the BSG nanostructure be created uniform.

The proposed formation mechanism enables fabrication of nanochannels not only with different transverse sizes (see pos. 2 and 3 in Fig. 9) but with a varying cross-section as well. To fabricate nanochannels having a varying cross-section, during the relative shift of layers, the cleaving front should be involved, besides the translational motion, in a slight rotational motion around the axis perpendicular to the small facet plane.

## 6. Discussion

The detected nanostructure has been formed as a result of a number of inelastic deformations. At the moment of the nanostructure formation, the ultimate relative elongation apparently approached the maximum permissible level for graphene (13% for the "armchair" orientation; 20% for the "zigzag" orientation) [11, 29, 30] or even exceeded it at some locations (see the structure defects in Fig. 1 in the form of ruptured upper facets). Immediately as the box-shaped nanostructure is being formed, the high stresses in its elements relax through inelastic mechanisms: sliding of the cleaving front, bending of the graphite/graphene plane layers, splitting of the layers in the folds, shifting of the split layers relative to each other, structural rearrangements in the graphene layer [30], and in the extreme case through a complete breakage of C–C bonds. In the absence of inelastic mechanisms, the whole steady spatial nanostructure, which we observe, could not have been formed since after the external force is removed, it would simply have returned to the initial state – a "stack" of graphene sheets.

### 6.1. Single crystal graphite versus HOPG

As shown above, the cleaving force should be oriented relative to graphite crystal lattice in such a way that the cleaving front be parallel to any of the three crystallographic directions of the basal plane. The easiest way to maintain a certain orientation of the cleaving front is to use single crystal graphite (SCG) or Kish graphite [31, 32] instead of HOPG. The point is that HOPG macrosample is a polycrystal, where the normals to the basal planes ($c$ directions) of all the crystallites are oriented nearly along the same direction (mosaic spread is tenths of a degree) and other directions ($a$ and $b$) of the crystallites are randomly oriented. Therefore, with HOPG, it is impossible to immediately set the required orientation of the cleaving front relative to crystallographic directions and so we should only rely on a chance that somewhere at the surface a crystallite exists with the necessary orientation. Therefore, in case HOPG is applied, in order to detect the sought for box-shaped nanostructure, the entire cleavage area has to be looked through. Actually a search should be performed for a crystallite that satisfies the above condition of cleaving front orientation. This conclusion means that the described method of box-shaped nanostructure fabrication on HOPG is rather time-consuming in the sense of searching for the prepared nanostructure itself.

The rare character of spontaneous formation of the BSG nanostructure while HOPG cleaving is confirmed by the circumstances of the nanostructure discovery. The box-shaped nanostructure was first found during trials and refinements of the method of distributed calibration of a probe microscope scanner [25] based on FOS approach [23, 24]. During these works, the overall measurement time made up more than a year of continuous scanning in automatic mode. Approximately once a day, an overview 2×2 μm scan was carried out at a new location of the sample. It is worth to note that while conducting the distributed calibrations, the HOPG sample was cleft, in fact, as rarely as about once in 2-3 months [25]. Interestingly, the most of the structures previously published in scientific literature, including various superlattices [33, 34, 35], were observed on the HOPG surface when the measurements were being taken.

### 6.2. Moiré superlattices inside nanochannels

Considering that the box-shaped nanostructure is formed as a result of a relative shift of the graphene layers, one may suppose that moiré superlattices [33, 34, 35] are likely to appear in the contact area of these layers (see Fig. 10). Although the shift of the layers in BSG nanostructure occurs mostly in parallel to a graphite crystallographic direction, the formed moiré pattern will not necessarily be a system of 1D fringes. Since the cleaving front

R. V. Lapshin

oriented at the angle $\alpha$ is not strictly parallel to the crystallographic direction $\theta_3$ of the upper graphene layer (see Section 4, Fig. 7), the graphene layers forming the box-shaped nanostructure may have rotated relative to each other at an angle of order $\alpha$-$\theta_3$=4.7°. Such angles are sufficient enough for formation of a 2D hexagonal superlattice having a period of several nanometers and corrugations up to 2 nm [33, 34, 35].

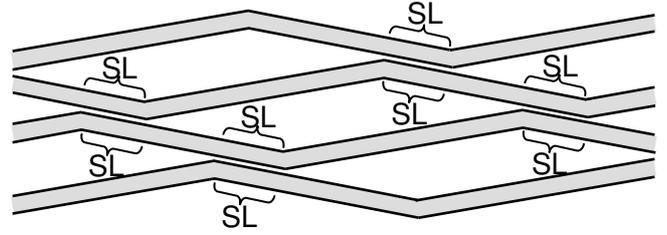

Fig. 10. Locations of the possible moiré superlattice formation (designated with the letters SL) on the inner surface of the channels of the box-shaped nanostructure.

Note that unlike the hexagonal moiré pattern for which formation it is sufficient to rotate one graphene layer relative to the other one, in order for a pattern of moiré fringes to be formed, it is required that one graphene layer be stretched/compressed relative to the other layer. Deformation of the contacting graphene layers simultaneously along *x* and *y* directions also leads to the formation of a 2D centered-hexagonal superlattice. Since tensile/compressive strains can only be small from the physical point of view [29, 30], they cause moiré patterns with large periods. By changing the parameters of the graphene superlattices formed on the inner surface of the channels of the box-shaped nanostructure, it is possible to modify the energy spectrum of electrons in these areas [36] as well as to control the adsorption properties of the nanochannels [34]. Moreover, the hexagonal superlattices can be used as a template for making ordered nanostructures [37] on the inner surface of the nanochannels.

*6.3. Additional methods to control the formation of BSG nanostructure*

Although the formation of the nanostructure considered is a pure result of a random unmanaged process, a number of factors apparently predetermined its appearance. Among the factors are: the specific orientation of the cleavage front relative to crystallographic directions of the graphite basal plane, the specific value of a cleaving force and the specific relationship between the lateral and the vertical components of the cleaving force as well as the specific direction of the cleaving force relative to the basal plane.

Moreover, by implementing a certain pattern of mechanical stresses/defects on and/or near the graphite surface that weaken the bonds between the graphite planes in some surface areas and strengthen them in other areas, an attempt could be made to take a better control of the process of the box-shaped nanostructure formation (folding, layer splitting and shifting). In order to implement such stress/defect pattern, some of the already known physical and/or chemical methods could be applied. Among them are electron/ion bombardment [38, 39], intercalation [32], substrate thermal deformation [40, 41], surface "cutting" by means of catalytic hydrogenation [42] or local probe oxidation [43], *etc*.

*6.4. Covering inner surface of nanochannels*

The suggested mechanism of BSG nanostructure formation also implies that if HOPG is able to intercalate [32] some substance into the surface layer then it is possible, if necessary, to cover (modify) the internal surface of the channels of the box-shaped nanostructure with an atomic layer of that substance. At the first stage, after the BSG nanostructure formation, the covering appears at least on two walls of the channel. At the second stage, the covering material is transferred onto the other two walls by means of annealing in vacuum. Intercalation of the atoms that form a dielectric layer allows fabrication of nanochennels with upper and lower parts isolated from each other so the nanochennels can be used as electrodes (for example, to apply a transverse electric field [44]).

# STM observation of a box-shaped graphene nanostructure

*6.5. Possible applications*

The practical importance of the discovered phenomenon consists in the fact that rather complicated multilayer hollow 3D nanostructures of graphene do exist in principle and that they can be fabricated by using original graphite as a raw stock. It is well known that graphene is especially worthwhile being a thin (literally atomic) graphite layer completely separated from a substrate [16, 45]. Otherwise, this material degenerates into regular, yet very thin, carbon film, which per se can be easily fabricated by the contemporary well-developed methods of molecular-beam epitaxy (MBE) [46, 47] or chemical vapor deposition (CVD) [39, 48].

What is important for practical application of the detected nanostructure is that the cross-section of the formed channels can vary widely. Unlike the nanopores existing in graphene [49], the nanopores (nanochannels) of BSG nanostructure are not perpendicular to the basal plane but are parallel to it. Moreover, the edge of the open nanopore (see Fig. 1) is so sharp that it is able to "resolve", for instance, single nucleotide bases in a DNA molecule while translocating through the nanopore [49].

In perspective, BSG nanostructures can be used to create ultra-sensitive detectors [14, 16], high-performance catalytic cells, nanochannels of microfluidic devices (molecular sieving, DNA sequencing and manipulation) [44, 49, 50, 51], high-performance heat sinking surfaces, rechargeable batteries of enhanced performance [50], nanomechanical resonators [14, 15, 16, 17], electron multiplication channels in emission nanoelectronic devices, high-capacity sorbents for safe hydrogen storage [52].

## 7. Conclusions

The key points of the research can be summarized as follows:

(1) A previously unknown 3D box-shaped graphene nanostructure has been detected on highly oriented pyrolytic graphite after mechanical cleavage.

(2) The discovered nanostructure is a multilayer system of parallel hollow nanochannels having quadrangular cross-section with typical width of a nanochannel facet 25 nm, typical wall/facet thickness 1 nm and length 390 nm and more.

(3) An original mechanism has been proposed that qualitatively explains the formation of the nanostructure detected. To elaborate a more detailed mechanism, a more detailed investigation of the nanostructure is required including computer modeling and attempts of intentional fabrication.

(4) Applications are revealed where the use of the box-shaped graphene nanostructure may lead to new scientific results or improve performance of existing devices.

**Acknowledgments**

This work was supported by the Russian Foundation for Basic Research (project 16-08-00036) and by the Ministry of Education and Science of the Russian Federation (contracts 14.429.11.0002, 14.578.21.0009). I thank O.E. Lyapin for critical reading of the manuscript and discussions; Dr. O.V. Sinitsyna for discussion on dislocation networks; Dr. A.L. Gudkov, Prof. E.A. Ilyichev and Assoc. Prof. E.A. Fetisov for their support and encouragement.

# STM observation of a box-shaped graphene nanostructure